\documentstyle[12pt]{article}
\nopagebreak

\begin{document}

\centerline{\Large\bf Asymmetry of the missing momentum
distribution}
\centerline{\Large\bf in (e,e$'$p) reactions and color
transparency}

\vskip 1.2truecm

\centerline{{\bf A. Bianconi, S. Boffi and D.E. Kharzeev}
\footnote{On leave from Moscow University, Moscow, Russian
Federation}}
\medskip

\centerline{Dipartimento di Fisica Nucleare e Teorica,
Universit\`a di Pavia, and}

\centerline{Istituto Nazionale di Fisica Nucleare,
Sezione di Pavia, Pavia, Italy}

\vskip 12pt


\centerline{\bf Abstract}

\medskip

We suggest the measurement of the integrated asymmetry
of the missing momentum distribution in (e,e$'$p) reactions
to check color transparency effects at intermediate
momentum transfers.

\vskip 1.2truecm


Exclusive (e,e$'$p) experiments are considered as one
of the best tools to study the phenomenon of Color
Transparency (CT) (see, e.g., [1,2] and references therein).
Within various models
it has been shown [3-7] that in (e,e$'$p)
reactions a complete transparency can be expected
at momenta  of the ejected nucleon $p'\sim 20$ GeV or
more. This region is
at present far from experimental possibilities.
Present experiments [see, e.g., [8]) can however explore the
region $p' \sim 2  \div 10$ GeV where a nontrivial and
irregular behaviour of CT is foreseen.
It has been recently shown that Fermi motion plays
a crucial role in the onset of CT [4,5]. In particular,
at intermediate energies CT is strongly affected by the
longitudinal component of the missing momentum of the
reaction [6,7]:
$p_m$ $\equiv$ $( \vec {p}\, ' - \vec q ) \cdot (\vec q /q)$,
where $\vec q$ is the momentum transferred by the virtual
photon.
Measurements of nuclear transparency as a function of
both $q$ and $p_m$ should, in principle,
afford seeing CT effects at relatively
low energies.

However, three problems are present:

1) At $p,q \sim 5 \div 10$ GeV it is difficult to measure
$ p_m$ with good precision, taking into account
that the useful values of $p_m$ range from $-200$ to
200 MeV.

2) CT effects must be disentangled from nuclear effects.
Traditionally, in the theoretical works this is accomplished
by defining a transparency coefficient $T$ as the ratio
between the calculated yield and a PWBA evaluation.
But the real number of events as a function of $p_m$ is
largely dominated by the PWBA distribution, which is
strongly peaked and changes by at least one order of magnitude
in the range $0\le p_m\le 200$ MeV. It is hard to determine
experimentally the ratio $T$ in the tails of the momentum
distribution, where the rate of events is low. In addition,
one needs to divide the number of events by a PWBA
calculation, which is model dependent.

3) On the experimental side one usually defines transparency
as the ratio of the measured nuclear cross section to the
summed elementary cross section of $Z$ (incoherent) protons.
This is useful when applied to integrated cross sections,
but cannot give too much information on anomalous transparency
when looking at a $p_m$ distribution. Indeed, such a
distribution is dominated by the PWBA behaviour. So plotting
such a ratio one would just more or less see a PWBA shape.

A possible way out of these problems is to remember that
the PWBA distribution is symmetric under the exchange
$p_m \rightarrow -p_m$, when $\vec q$ and the transverse
component $p_t$ of the missing momentum are kept fixed.
At high momenta even the DWBA calculation performed in the
Glauber formalism approximately shows the same symmetry, if
one neglects any CT effects. On the contrary, CT effects are
largely dependent
on the sign of $p_m$ [6,7]. The reason is the following:
if it exists, CT is due to the formation of a compact
(ideally {\sl pointlike}\/) baryonic state when the virtual
photon is
absorbed by a proton in the nucleus. This compact
state cannot be of course a proton on its mass-shell, but
it is a superposition of many baryon states with different
mass: the proton, its resonances,
the continuum. It is well known that the threshold for the
production of each of these states depends on
$p_m$ in an {\sl asymmetric}\/ fashion. As an
example we show in fig. 1 the loci of maximum probability for
excitation of some intermediate states with given mass
as a function of $q$ and $p_m$ (see also ref. [7]).
It is evident that all the
excited baryon states are produced preferentially at positive
$p_m$. So a compact state is better realized at positive
$p_m$, since one can {\sl simultaneously}\/
produce both a real proton and states with $m_* > m_p$.

Let us define $N_+$ ($N_-$) as the number of
events in the region $x< p_m <y$ ($-y< p_m < -x$)
at given $q$ and $p_t$. The threshold
$x$ can be zero, or some tens of MeV; $y\sim 100\div 200$ MeV.
We suggest to measure the asymmetry

$$A \equiv\ {{N_+ - N_-}\over {N_+ + N_-}}.$$

The values of $A$ calculated in parallel kinematics ($p_t =0$)
as a function of $q$ at different
$x$ are shown in figs. 2 and 3 at $y=100$ and $200$ MeV,
respectively. The adopted model is the same as in refs. [4,6],
improved by the three-channel calculation [7].

The reason to have a central window ($-x,x$) of lost
events is to exclude the region where the symmetric PWBA
peak dominates the distribution. As one can clearly see
in figs. 2 and 3 the larger is $x$ the larger is $A$. At the
same time, a large value of $x$ restricts the statistics to
a very small number of events. An optimal choice depends on
the values of the fixed variables, $q$ and $p_t$, and on the
experimental acceptances.

An upper threshold between 100 and 200 MeV is needed to
exclude those regions (in the tails of the nuclear Fermi
distribution) where the validity of impulse approximation
may be questionable and an asymmetry might arise from other
mechanisms  than CT.

The clear advantages of this measurement are:

1) One is not restricted to measure the missing momentum
$p_m$ with a good precision: it is possible to choose just
two wide $p_m$ regions (e.g.: $x = 0, y = 150$ MeV).
Ideally, the proposed measurement is to be performed
in parallel kinematics. However, concerning the transverse
component $p_t$ one can always choose
either a given $p_t$, or a wide $p_t$ region to integrate
it over.

2) The normalization is model-independent.

The estimated magnitude of the asymmetry
(figs. 2 and 3) hopefully should allow to observe it
in present or coming experiments. However, in the
experimental conditions of current NE18 experiment
at SLAC [8] the kinematics is close to perpendicular
and only a very limited range of $p_m$ can be explored [8,9].
Therefore we think that specially dedicated experiments
are desirable.

We would like to stress that the origin of
the asymmetry in the missing momentum distribution
is largely model-independent. Its appearance
only requires that the compact state is a superposition
of different mass eigenstates with similar couplings.
This is a straightforward consequence of the foundations
of the CT theory.

\medskip

\vskip 1.2truecm

We are grateful to R.D. McKeown and R.G. Milner for useful
discussions.

\vfill

\clearpage
\centerline{\Large\bf Figure captions}
\vskip 0.5cm
Fig. 1. The maxima of the production probability of an
intermediate state with given mass $m_*$ for the
$^{12}$C(e,e$'$p) reaction as a function of three--momentum
transfer $q$ (GeV) and longitudinal missing momentum $p_m$
(in units of Fermi momentum $p_F=221$ MeV).
Solid (dotted) line for $m_*= 1.44$ (1.8) GeV.

Fig. 2. The asymmetry of the number of events in the
$^{12}$C(e,e$'$p) reaction as a function of three--momentum
transfer $q$ (GeV) integrated over the longitudinal missing
momentum
$p_m$ in the region $x\le p_m\le y = 100$ MeV. Solid, dashed
and dotted lines refer to $x= 10, 30, 50$ MeV, respectively.

Fig. 3. The same as in fig. 2, but with $y = 200$ MeV.

\end{document}